\documentclass[a4paper,twoside]{article}

\usepackage{placeins}
\usepackage{epsfig}
\usepackage{subfigure}
\usepackage{calc}
\usepackage{amssymb}
\usepackage{amstext}
\usepackage{amsmath}
\usepackage{amsthm}
\usepackage{multicol}
\usepackage{pslatex}
\usepackage{apalike}
\usepackage{color}
\usepackage{listings}
\usepackage{tikz,ifthen}
\usepackage{hyperref}
\usepackage{paralist}
\usepackage{xspace}
\usepackage{algorithm}
\usepackage{algorithmic}

\usepackage{SCITEPRESS}     % Please add other packages that you may need BEFORE the SCITEPRESS.sty package.

\usepackage{tikz,ifthen,pgfplots}
\usetikzlibrary{arrows,trees,backgrounds,automata,shapes,decorations,plotmarks,fit,calc,positioning,shadows,chains}
\tikzstyle{every pin edge}=[<-,shorten <=1pt]
\tikzstyle{neuron}=[circle,fill=black!25,minimum size=17pt,inner sep=0pt]
\tikzstyle{input neuron}=[neuron, fill=green!50]
\tikzstyle{output neuron}=[neuron, fill=red!50]
\tikzstyle{hidden neuron}=[neuron, fill=blue!50]
\tikzstyle{annot} = [text width=6em, text centered]
\usetikzlibrary{calc}

\DeclareFixedFont{\ttb}{T1}{txtt}{bx}{n}{7.5} % for bold
\DeclareFixedFont{\ttm}{T1}{txtt}{m}{n}{7.5}  % for normal

\definecolor{deepblue}{rgb}{0,0,0.5}
\definecolor{deepred}{rgb}{0.6,0,0}
\definecolor{deepgreen}{rgb}{0,0.5,0}

\newtheorem{lemma}{Lemma}

\newcommand{\ra}{\rangle}

\newcommand{\compose}{\parallel}

\newcommand{\request}{{\color{blue}request}}
\newcommand{\waitfor}{{\color{green!50!black}wait for}}
\newcommand{\blocking}{{\color{red}blocking}}

\begin{document}

\title{Towards Repairing Scenario-Based Models with Rich
  Events}

\author{\authorname{Guy Katz}
 \affiliation{The Hebrew University of Jerusalem, Jerusalem, Israel}
 \email{guykatz@cs.huji.ac.il}}

\keywords{Scenario-Based Modeling, Program Repair, Model Checking,
  Constraint Solvers, SMT Solvers}

\abstract{Repairing legacy systems is a difficult and error-prone
  task: often, limited knowledge of the intricacies of these systems
  could make an attempted repair result in new errors. Consequently,
  it is desirable to repair such systems in an automated and sound
  way. Here, we discuss our ongoing work on the automated repair of
  \emph{scenario-based models}: fully executable models that describe
  a system using \emph{scenario objects} that model its individual
  behaviors. We show how rich, scenario-based models can be
  model-checked, and then repaired to prevent various safety
  violations. The actual repair is performed by adding new scenario
  objects to the model, and without altering existing ones --- in a
  way that is well aligned with the principles of scenario-based
  modeling. In order to automate our repair approach, we leverage
  off-the-shelf SMT solvers. We describe the main principles of our
  approach, and discuss our plans for future work.}

\onecolumn \maketitle \normalsize \vfill

\section{\uppercase{Introduction}}
\label{sec:introduction}
\noindent

Modeling complex systems is a painstaking and difficult task. Even
once a suitable model has been created, and the system in question has
been implemented and deployed, the model may still need to be changed
as part of the system's life cycle --- for example, if bugs are
discovered, or if the specification of the system is 
changed. This post-deployment altering of systems and models, which we
refer to as \emph{repair}, is a challenging undertaking: even if the
desired change is small, i.e. if it only affects a small portion of
the system's operations, attempting a fix could have undesirable
consequences. For example, dependencies between various system
components, which have not been property modeled or documented, could
make changing one component affect other components in unintended
ways. This problem is typically compounded by lack of knowledge ---
because the engineers who developed the system are
unavailable, or have forgotten crucial details. Thus,
as program repair is frequently needed, we require formalisms and
tools that will allow us to \emph{automatically} repair systems and
models in a safe and convenient way.

One promising approach for tackling this difficulty is through
modeling techniques that facilitate model repair.
\emph{Scenario-Based Modeling}
(\emph{SBM})~\cite{DaHa01,HaMa03,HaMaWe12ACM} is a notable candidate
that fits this description. In SBM, systems are modeled through the
specification of \emph{scenario objects}: objects that represent
individual behaviors of the system being modeled. Each of these
objects describes either behavior that the system should uphold, or
behavior that it should avoid. Although each object is only tasked
with governing a narrow aspect of the overall system behavior, the
resulting model is fully executable --- i.e., the various objects can
be composed together and executed, in a way that achieves the overall
system goals. This execution is performed by an \emph{event selection
  mechanism}, which is in charge of executing the objects
simultaneously and synchronizing them in a way that produces cohesive
behavior. Studies have shown that SBM is nicely aligned with how
humans perceive systems, and that it consequently fosters abstract
programming~\cite{GoMaMe12Spaghetti,AlArGoHa14}.

Although SBM was designed as a general modeling framework, not
particularly geared towards model repair, prior work has shown its
compatibility with various formal analysis techniques, such as
model-checking~\cite{HaLaMaWe11} and automated
repair~\cite{HaKaMaWe14}. This compatibility stems from the fact that
scenario objects in a scenario-based (SB) model interact through the
well defined interface of the event selection mechanism, making it
possible to automatically construct a model of the full, composite
system, and then analyze it. Tools and techniques have been devised to
model-check and repair safely and liveness violations in SB
models~\cite{HaLaMaWe11,HaLaMaWe11}, to automatically optimize and
distribute these models~\cite{HaKaKa13,HaKaKaMaWeWi15,StGrGrHaKaMa17},
and to identify emergent properties thereof~\cite{HaKaMaMa18}.

Recently, it has been observed that while SBM is well equipped for
modeling reactive systems, it is sometimes inadequate for modeling
systems that handle data --- for example, robotics and autonomous
vehicle systems~\cite{KaMaSaWe19,Ka20}, which involve various
mathematical computations in addition to their reactivity.  To this
end, researchers have extended the SBM principles, allowing the event
selection mechanism to support \emph{rich events}, i.e. events that
carry various types of data.  These enhancements have proven quite
useful for modeling more complex systems, but are unfortunately
incompatible with existing formal analysis and repair techniques for
SBM, which rely on the simplistic nature of the event selection
mechanism. This has raised the following question: \emph{are the
  automated analysis benefits afforded by SBM limited to the simple
  models, or do these carry over when SBM is used in richer settings?}

Here, we begin to answer this question, by devising analysis
techniques for rich SBM. Specifically, we focus
on program repair: we show how, using appropriate extensions, the
automated repair techniques proposed for SBM carry over when richer
events are introduced and more complex systems are modeled. At the
core of our proposed extension is the ability to automatically extract
the underlying transition graphs of SB models with rich events,
through the use of SMT solvers --- a powerful family of automated
solvers that can reason about first order logic theories, such as
arithmetic. We propose an SMT-based method for constructing SBM
transition graphs, which effectively reduces sets of infinitely many
possible events into a finite set of possibilities that need to be
explored. Although our work is focused primarily on program repair, we
believe it will allow the extension of other automated analysis
techniques to the rich SBM setting.

The rest of this paper is organized as follows. In
Section~\ref{sec:background} we provide the necessary background on
SBM, and its extension to handle rich events. Next, in
Section~\ref{sec:transitionGraphs} we present our core technique for
automatically extracting the underlying transition graphs of SB models
with rich events. We then show how this technique enables us to
automatically model-check and repair SB models with rich events in
Section~\ref{sec:repairSbm}. We discuss related work in
Section~\ref{sec:relatedWork}, and conclude in
Section~\ref{sec:conclusion}.

\section{\uppercase{Background}}
\label{sec:background}
\noindent

\subsection{Vanilla Scenario-Based Modeling}
The multiple variants of SBM that have been proposed typically include
a set of scenario objects that are run in parallel, and repeatedly
synchronize with each other.  the most commonly used synchronization
idioms~\cite{HaMaWe12ACM} include:
\begin{enumerate}
\item \emph{requesting events}: a scenario object can \emph{request} event
  $e$, indicating that it wishes that $e$ be triggered. Intuitively, $e$
  represents some desirable behavior that the system should now
  perform.
\item \emph{waiting-for events}: a scenario object may \emph{wait-for} an
  event, i.e. state that it wishes to be notified when that event is
  triggered. However, the object does not actively request this
  event. The waiting-for idiom is useful, for example, when an object
  is waiting for some sequence of external events to mark that it
  should perform some actions in response.
\item \emph{blocking events}: when a scenario object \emph{blocks}
  event $e$, it prevents the overall system from triggering it ---
  even if $e$ was requested by another object. Intuitively, this is
  how components can indicate undesirable behavior, which the
  system is forbidden from performing.
\end{enumerate}
We refer to the SBM variant that consists of these idioms as
\emph{vanilla SBM}, or \emph{vSBM}, for short. in vSBM, each scenario
object defines a set of synchronization points, and in each of these
points it declares its sets of  requested, waited-for and
blocked events. The event selection mechanism collects these declarations
from all objects; selects for triggering one \emph{enabled event,}
i.e. an event that is requested
and not blocked; and then informs all the objects that requested or
waited-for this event about the selection. These objects then progress
to their next synchronization point, and the process is repeated as
the execution continues.

\begin{figure}[htp]
  \centering
  \scalebox{0.65} {
    
    \tikzstyle{box}=[draw,  text width=2cm,text centered,inner sep=3]
    \tikzstyle{set}=[text centered, text width = 10em]

    \begin{tikzpicture}[thick,auto,>=latex',line/.style ={draw, thick, -latex', shorten >=0pt}]
      
      \matrix(bts) [row sep=0.3cm,column sep=2cm]  {

        \node (box1)  [box] {\waitfor{} \textsc{WaterLow}}; \\
        \node (box2)  [box] {\request\ \textsc{AddHot}}; \\
        \node (box3)  [box] {\request\ \textsc{AddHot}}; \\ 
        \node (box4)  [box] {\request\ \textsc{AddHot}}; \\ 
      };

      \draw [->] ($(box1.north) + (0,0.3cm)$) -- (box1.north);
      \node (title) [above=0.1cm of bts,box,draw=none] at ($(bts) + (-0.25cm,2.51cm)$) 
      {\textsc{AddHotWater}};  
      
      \begin{scope}[every path/.style=line]
        \path (box1)   -- (box2);
        \path (box2)   -- (box3);
        \path (box3)   -- (box4);
        \path (box4.east)   -- +(.25,0) |- (box1);
      \end{scope}

      \matrix(bts2) [right=.25cm of bts, row sep=0.3cm,column sep=2cm] {
        \node (box1)  [box] {\waitfor{} \textsc{WaterLow}}; \\
        \node (box2)  [box] {\request\ \textsc{AddCold}}; \\
        \node (box3)  [box] {\request\ \textsc{AddCold}}; \\ 
        \node (box4)  [box] {\request\ \textsc{AddCold}}; \\ 
      };
      
      \draw [->] ($(box1.north) + (0,0.3cm)$) -- (box1.north);
      \node (title) [above=0.1cm of bts2,box,draw=none] at ($(bts2) + (-0.25cm,2.51cm)$) 
      {\textsc{AddColdWater}};

      \begin{scope}[every path/.style=line]
        \path (box1)   -- (box2);
        \path (box2)   -- (box3);
        \path (box3)   -- (box4);
        \path (box4.east)   -- +(.25,0) |- (box1);
      \end{scope}

      \matrix(bts3) [right=.25cm of bts2, row sep=0.3cm,column sep=2cm] {
        \node (box1)  [box] {\waitfor{}  \textsc{AddHot} while  \blocking\ \textsc{AddCold}}; \\
        \node (box2)  [box] {\waitfor{}  \textsc{AddCold} while \blocking\ \textsc{AddHot}}; \\
      };

      \draw [->] ($(box1.north) + (0,0.3cm)$) -- (box1.north);
      \node (title) at (title-|bts3) [box,draw=none] {\textsc{Stability}};  

      \begin{scope}[every path/.style=line]
        \path (box1)   -- (box2);
        \path (box2.east)   -- +(.25,0) |- (box1);
      \end{scope}
      
      \node (log)  [right=.3cm of bts3,box,text width=2cm,fill=yellow!20] {
        $\cdots$ \\ 
        \textsc{WaterLow} \\
        \textsc{AddHot}  \\ 
        \textsc{AddCold} \\ 
        \textsc{AddHot}  \\ 
        \textsc{AddCold} \\ 
        \textsc{AddHot}  \\ 
        \textsc{AddCold} \\ 
        $\cdots$ \\
      }; 

      \node (title2) at (title-|log)            
      [box,draw=none] {\textsc{Event Log}};  
    \end{tikzpicture}
  }  
  \caption{
    (From~\cite{HaKaMaWe14})
    A scenario-based model of a
    system that controls the water level in a tank with hot and
    cold water taps. 
  }  
  \label{fig:watertap}
\end{figure}
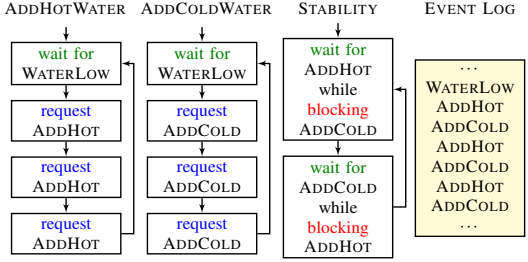

A toy example, borrowed from~\cite{HaKaMaWe14}, is depicted in
Fig.~\ref{fig:watertap}. The model that appears therein belongs to a
system that controls the water level in a tank. Adding water can be
done from either a hot water tap or a cold water tap. The various
scenario objects are each depicted as a transition system: the nodes
represent the object's synchronization points, and they are labeled
with the events that are requested, waited-for and blocked in those
points. The scenario object \textsc{AddHotWater} repeatedly waits for
\textsc{WaterLow} events and requests three times the event
\textsc{AddHot}; and the scenario object \textsc{AddColdWater}
performs a symmetrical operation with cold water. If the model only
includes these two objects, \textsc{AddHotWater} and
\textsc{AddColdWater}, the three \textsc{AddHot} events and three
\textsc{AddCold} events may be triggered in any order when the model
is executed. If, for example, we wish to maintain a steady water
temperature, we may add the scenario object \textsc{Stability} in
order to enforce the interleaving of \textsc{AddHot} and
\textsc{AddCold} events, by using event blocking. The execution trace
of the resulting model (with all three objects) is depicted in the
event log.

We follow the definitions of~\cite{Ka13}, and formalize the
vSBM framework as follows. Given some finite event set
$E$, we define a scenario object $O$ as the tuple
$O = \langle Q, \delta, q_0, R, B \ra$, where the interpretation of
the components is as follows:
\begin{itemize}
\item $Q$ is a set of states. Each state represents a
  predetermined synchronization point.
\item $q_0\in Q$ is the initial state.
\item $R$ and $B$ are mappings, $R,B:Q\to 2^E$. They map states to set
  of events requested ($R$) and blocked ($B$) at those states. 
\item $\delta: Q \times E \to 2^Q$ is a transition function. It 
  indicates how the object reacts when an event is triggered, i.e. if
  $q'\in\delta (q,e)$ then the object can transition to state $q'$
  when event $e$ is triggered in state $q$.
  \end{itemize}
We sometimes refer to this tuple as the \emph{underlying transition
  graph} of scenario object $O$.
  
  The composite model specified by a set of scenario objects is
  defined using a composition operator, which combines two scenario objects
  into a single, larger scenario object, as follows.  For two scenario
  objects $O^1 = \langle Q^1, \delta^1, q_0^1, R^1, B^1 \ra$ and
  $O^2 = \langle Q^2, \delta^2, q_0^2, R^2, B^2 \ra$, both over a
  common event set $E$, we define the composite scenario object
  $O^1\compose O^2$ as
  $O^1 \compose O^2 = \langle Q^1 \times Q^2, \delta, \langle
  q_0^1,q_0^2\rangle, R^1\cup R^2, B^1\cup B^2\rangle $, where:
\begin{itemize}
\item The transition relation is defined element-wise, i.e. $\langle \tilde{q}^1,\tilde{q}^2\rangle \in  \delta(\langle q^1,q^2\rangle, e)$
if and only if $\tilde{q}^1 \in \delta^1(q^1,e)$ and $\tilde{q}^2\in
\delta^2(q^2,e)$.
\item The labeling of a composite state is the union of the
  element-wise labeling, i.e.  $e\in (R^1\cup
R^2)(\langle q^1,q^2 \rangle)$ if and only if $e \in R^1(q^1) \cup
R^2(q^2)$, and
$e\in (B^1\cup
B^2)(\langle q^1,q^2 \rangle)$ if and only if $e \in B^1(q^1) \cup
B^2(q^2)$.
\end{itemize}

Finally, we define a \emph{behavioral model} $M$ as a collection of
scenario objects $O^1, O^2,\ldots, O^n$. The executions of $M$ are
then defined to be the executions of the composite scenario object
$O = O^1\compose O^2\compose\ldots\compose O^n$.  Each execution of
$M$ starts from the initial state of $O$, and in each state $q$ along
the run it selects for triggering an enabled event, i.e., an event
$e\in R(q) - B(q)$ (if no such event exists, the execution terminates
in a deadlock).  Then, the execution moves to a state
$\tilde{q}\in \delta(q,e)$, and so on.

In practice, users very seldom describe models by providing the
transition graphs of their scenario objects.  Instead, SBM has been
implemented in a variety of tools, either as dedicated frameworks
(e.g., the Play-Engine tool for Live Sequence Charts
(LSC)~\cite{HaMa03} or the ScenarioTools~\cite{GeGrGuKoGlMaKa17}
engine), or on top of popular programming languages, such as
JavsScript~\cite{BaWeRe18}, Python~\cite{bppy}, Java~\cite{HaMaWe10BPJ} and
C++~\cite{HaKa14}. SBM has been used in modeling complex systems, such
as robotic controllers~\cite{ElSaWeYa19,GrGr18b},
web-servers~\cite{HaKa14}, smart buildings~\cite{ElMaStWe18},
a nano-satellite~\cite{BaElSaWe19}, and
cache coherence protocols~\cite{HaKaMaMa16}.

\subsection{Scenario-Based Modeling with Rich Events}

Although vanilla SBM has been successfully used in various contexts,
in recent years it was shown that it may fall short in expressing
various complex interactions between scenario
objects~\cite{KaMaSaWe19,Ka20,El20}. Specifically, the simple event
declaration mechanism --- a finite set of events $E$, and a finite set
of requested, waited-for and blocked events in every state, may be
inadequate for expressing more complex behaviors.

For example, consider a drone that needs to turn left or right by a
certain degree. Degrees are represented as real numbers, and an
attempt to express this using vanilla SBM would require either some loss
of precision, e.g. by discretizing the set of possible degrees; or the
use of various hacks that circumvent the SBM event selection mechanism,
thus going against the grain of SBM.

In line with the notions in~\cite{KaMaSaWe19}, we define \emph{rich
  SBM} (\emph{rSBM}) as follows: instead of discrete events, the event
set $E$ now contains a set of real-valued variables. Each scenario
object can now request that certain variables in $E$ be assigned
certain values, and block certain variables from being assigned other
values. Event selection then consists of choosing a variable
assignment that satisfies at least one request, and violates none of
the blocked assignments.

\begin{figure}[htp]
  \centering
  \scalebox{0.65} {
    
    \tikzstyle{box}=[draw,  text width=2.8cm,text centered,inner sep=3]
    \tikzstyle{set}=[text centered, text width = 10em]

    \begin{tikzpicture}[thick,auto,>=latex',line/.style ={draw, thick, -latex', shorten >=0pt}]

      \node (box1)  [box, fill=white]
      {
        {\color{red}Block}\\$h\leq -20 \vee h\geq 20$
      };
      \path[] (box1) edge [loop below,thick] node {true} (box1);

      \node (box2)  [box, fill=white, below = 3cm of box1]
      {
        {\color{red}Block}\\$v\leq -5 \vee v\geq 5$
      };

      \path[] (box2) edge [loop below,thick] node {true} (box2);

      \node (box3)  [box, fill=white, right = 1cm of box1]
      {
        {\color{blue}Request} $v\geq 2 \wedge h=0$
      };

      \node (box4)  [box, fill=white, below = 1cm of box3]
      {
        {\color{blue}Request} $h\geq 10$\\
        {\color{red}Block} $v\neq 0\vee h<0$
      };
      \draw[->] (box3) -- node[] {$v\geq 2\wedge h=0$} (box4);

      \draw [->] ($(box1.north) + (0,0.5cm)$) -- (box1.north);
      \draw [->] ($(box2.north) + (0,0.5cm)$) -- (box2.north);
      \draw [->] ($(box3.north) + (0,0.5cm)$) -- (box3.north);

      \node (box5)  [box, fill=white, below = 1.5cm of box4]
      {
        {\color{blue}Request} true
      };
      \draw[->] (box4) -- node[] {$h\geq 10$} (box5);

      \node (box6)  [box, fill=white] at ($(box5.east) + (2cm, 1cm)$)
      {
        {\color{blue}Request} $h\geq 10$\\
        {\color{red}Block}\\$h<10\vee v\neq 0$
      };
      \path[->] (box4.east) edge [bend left] node[] {$h< 10$} (box6.north);

      \path[->] (box6.south) edge [bend left] node[] {$h\geq 10$} (box5.east);

      \path[] (box5) edge [loop below,thick] node {true} (box5);     
      
      \node (title) [box,draw=none] at ($(box1) + (-0.0cm,1.4cm)$) {\textsc{Object 1}};  
      \node (title) [box,draw=none] at ($(box2) + (-0.0cm,1.4cm)$) {\textsc{Object 2}};  
      \node (title) [box,draw=none] at ($(box3) + (+1.8cm,1.4cm)$) {\textsc{Object 3}};  

      \begin{pgfonlayer}{background}

      \draw[draw, fill=gray!10, rounded corners] ($(box1.south west) + (-0.2cm, -1.2cm)$)
      rectangle
      ($(box1.north east) + (0.2cm, 1.2cm)$);

      \draw[draw, fill=gray!10, rounded corners] ($(box2.south west) + (-0.2cm, -1.2cm)$)
      rectangle
      ($(box2.north east) + (0.2cm, 1.2cm)$);

      \draw[draw, fill=gray!10, rounded corners] ($(box5.south west) + (-0.2cm, -1.12cm)$)
      rectangle 
      ($(box3.north east) + (3.7cm, 1.2cm)$);

       \end{pgfonlayer}

    \end{tikzpicture}
  }  
  \caption{A model for controlling a drone, comprised of three
    scenario objects. Each state lists the requested and blocked
    assignment for that state; and the transitions are given as guard
    formulas, where a transition can be traversed if and only if the
    triggered assignment satisfies the guard.}
  \label{fig:rsbmExample}
\end{figure}
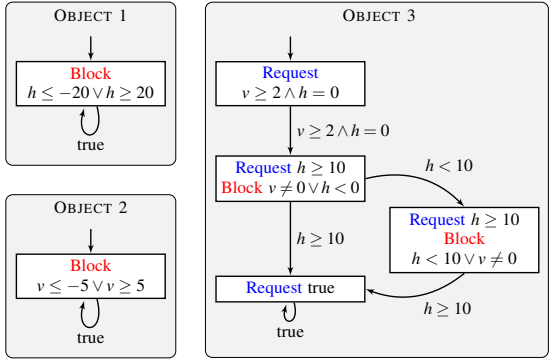

An example appears in Fig~\ref{fig:rsbmExample}. The model depicted
therein belongs to a system controlling a drone. The event set $E$
contains two real variables, $E=\{v,h\}$, representing the vertical
and horizontal angular velocities of the drone, respectively. Each
round of event selection assigns new values to these variables.
Scenario object 1 poses hard limits of $v$, due to mechanical
limitations in the drone; and scenario object 2 poses similar
constraints on $h$. Scenario object 3, in charge of navigating the
drone to its destination, requests an ascent at an angular velocity of
at least 2 degrees per second ($v\geq 2$), while not turning left or
right ($h=0$). Afterwards, it requests a right turn at 10 degrees per
second or more ($h\geq 10$), while blocking the drone from changing
altitude or turning left ($v\neq 0\vee h<0$). Depending on the actual
right turn that was performed, the object might request an additional
right turn at 10 degrees per second, this time forcing the drone to
turn by blocking all other possible assignments. Finally, the object
reaches its goal state, and requests nothing more.
  
Observe that the model in Fig.~\ref{fig:rsbmExample} again depicts the
scenario objects as transition systems. Further, the edges do not list
the (possibly infinite) set of assignments that trigger the
transition, but instead list \emph{guard formulas}: an edge may be
traversed only if the triggered assignment satisfies the edge's guard.

We formalize rich SBM as follows. The set $E$ is no longer a discrete
set of events, but is instead a set of real-valued
variables $E=\{x_1,\ldots,x_n\}$.  A scenario object $O$ is a tuple
$O = \langle Q, \delta, q_0, R, B \ra$, where $Q$ is again a set of
states and $q_0\in Q$ is an initial state. The labeling functions now
map each state $q\in Q$ into a first-order, linear real arithmetic
formula. Specifically, these formulas can impose linear constraints on
the variables e.g., $x_1\geq 5$ or $x_2+x_3\leq x_4$, and can have
arbitrary Boolean structure: e.g.,
$(x_1\geq 5)\rightarrow (x_2+x_3\leq x_4 \vee x_5 < 7)$. The
transition function $\delta:Q\times \mathbb{R}^n\to 2^Q$ now defines
how the state transitions for every possible assignment $\alpha$ that
assigns a real value to each of the variables $x_1,\ldots, x_n$.

For two scenario objects
$O^1 = \langle Q^1, \delta^1, q_0^1, R^1, B^1 \ra$ and
$O^2 = \langle Q^2, \delta^2, q_0^2, R^2, B^2 \ra$, both over a common
variable set $E$, we define the composite scenario object
$O^1\compose O^2$ as
$O^1 \compose O^2 = \langle Q^1 \times Q^2, \delta, \langle
q_0^1,q_0^2\rangle, R^1\vee R^2, B^1\vee B^2\rangle $. As before, the
transition relation is defined element-wise, i.e.
$\langle \tilde{q}^1,\tilde{q}^2\rangle \in \delta(\langle
q^1,q^2\rangle, e)$ if and only if $\tilde{q}^1 \in \delta^1(q^1,e)$
and $\tilde{q}^2\in \delta^2(q^2,e)$. The composite formulas that
represent the requested and blocked events are defined as the
disjunctions of the element-wise formulas:
$R^1\vee R^2(\langle q^1,q^2 \rangle) = R^1(q^1)\vee R^2(q^2)$ and
$B^1\vee B^2(\langle q^1,q^2 \rangle) = B^1(q^1)\vee B^2(q^2)$.

The \emph{rich behavioral model} $M$ is now defined as a collection of
rich scenario objects $O^1, O^2,\ldots, O^n$. The executions of $M$ are
then defined to be the executions of the composite scenario object
$O = O^1\compose O^2\compose\ldots\compose O^n$.  Each execution of
$M$ starts from the initial state of $O$, and in each state $q$ along
the run it selects for triggering a variable assignment $\alpha$, that
satisfies the formula
\[
  R(q)\wedge \neg B(q).
\]
In other words, the selected assignment satisfies the request of at
least one component object, and does not contradict the blocking
declarations issued by any object. 
 Then, the execution moves to a state
$\tilde{q}\in \delta(q,\alpha)$, and so on.

In practice, the discovery of an assignment that satisfies the given
constraints can be performed using various automated solvers, such as
LP or SMT solvers~\cite{HaKaMaSaWe20}. Because we restrict our
constraints to first-order, quantifier-free linear real arithmetic,
they can be resolved in polynomial time~\cite{BaTi18}.

\section{\uppercase{Formally Analyzing Rich Scenario-Based Models}}
\label{sec:transitionGraphs}
  
Much work has been put into performing formal, automated analysis of
SB models (e.g.,~\cite{HaLaMaWe11,Ka13,HaKaMaMa18}). The cornerstone
of these techniques is the automated extraction of the underlying
transition graph of an SB model $M=\{O^1,\ldots,O^n\}$ given in some
high-level language, such as C++. Intuitively, this is done in two
steps~\cite{Ka13}:
\begin{enumerate}
\item The underlying transition graph of each scenario object $O^i$ is
  extracted independently, directly from its code.  This process is
  performed by iteratively exploring the object's synchronization
  points. Starting at the initial state $q_0$, each non-blocked event $e\in E$ is
  triggered, and the object's reaction to $e$ is recorded.  If the
  object transitions to some state $q$, then the edge
  $q_0\stackrel{e}{\to} q$ is added to its transition graph. If $q$ is
  a previously unvisited state, it is added to a queue for later
  inspection. The process repeats until all possible events, in all
  reachable states, have been considered and mapped.
\item Once the transition graph for each scenario object has been
  extracted, these graphs are composed to produce the composite
  transition graph of $M$ (according to the composition operator
  defined in Section~\ref{sec:background}).
\end{enumerate}
A key point in this construction is that the event set $E$ is
finite. In particular, in Step 1 this allows us to exhaustively
trigger each non-blocked event in each state, and check how the object
transitions.  In order to apply a similar technique to rSB models,
this step needs to be adjusted; specifically, in an rSB model, the
discrete set of enabled events is replaced with a variable assignment,
and because there are infinitely many such assignments, enumerating
them is impossible. Step 2, on the other hand, remains unchanged also
when reasoning about rSBM, and can be applied to construct the
composite transition graph.

The method that we propose for extracting the transition graph from a
scenario object is as follows. We make the
observation that although there are infinitely many variable
assignments that the object needs to react to, these are
typically grouped into a finite number of possibilities that the
object handles in the same way. Consider, for example, the following
code snippet that defines a rich scenario object in some high-level
language:

\begin{lstlisting}
sync( $\color{blue} request$=($x< 5$) );
if ( $x \geq 2$ )
  A();
else
  B();
\end{lstlisting}

Here, the object synchronizes (using the \emph{sync} keyword) and
requests that $x$ be assigned a value less than $5$. Then, when the
synchronization call returns, i.e. when $x$ has been assigned a value
smaller than $5$, the object performs $A()$ if $x\geq 2$, and $B()$
otherwise. Thus, there is no difference between $x=3$ and $x=4$, as
far as the underlying transition graph is concerned; in either case,
the object will transition into the same state (the next
synchronization point in $A$). Additionally, the relevant predicates,
i.e. $(x\geq 2)$ in this case, are already available to us: we can
find them by simply parsing the code of the scenario object, and
collecting all the predicates that appear therein.

We thus propose the following approach. Given a scenario object $O$ in
some high-level language, we first parse its code and produce the set
$P_O$ of all predicates that appear in $O$ --- either as formulas
within its synchronization points, or elsewhere in the code. Next, at
every state $q$ of $O$, we observe the power-set $2^{P_O}$, and for
each element $\langle \varphi_1,\ldots,\varphi_k\rangle\in 2^{P_O}$ we use an SMT
solver to come up with a concrete assignment for which
$\varphi = \bigwedge_{i=1}^k \varphi_i$ holds. If such an assignment $\alpha$ exists, we
trigger it at state $q$, and record
\[
  q\stackrel{\varphi}{\to}q'
\]
in our transition graph.  The full algorithm appears as
Alg.~\ref{alg:spanTransitionGraph}.

%  objects $O^1 = \langle Q^1, \delta^1, q_0^1, R^1, B^1 \ra$ and
\begin{algorithm}
  \caption{Extract Transition Graph($O$)}
  \begin{algorithmic}[1]
    \STATE $P\leftarrow\ $ all predicates in $O$
    \STATE $Q.push( q_o )$
    \WHILE {$Q$ not empty}
      \STATE $q\leftarrow Q.pop()$
      \FOR {$\langle \varphi_1,\ldots,\varphi_k\rangle\in 2^{P_O}$}
        \STATE $\varphi\leftarrow \bigwedge_{i=1}^k\varphi_i$
        \STATE $\alpha\leftarrow $ SMT($\varphi$)
        \IF {$\alpha\neq\bot$}
          \STATE Invoke $\alpha$ in state $q$, mark new state as $q'$
          \STATE Add $q\stackrel{\varphi}{\to}q'$ to transition graph
          \IF {$q'$ not previously visited}
            \STATE $Q.push(q')$
          \ENDIF
        \ENDIF
      \ENDFOR
    \ENDWHILE
  \end{algorithmic}
  \label{alg:spanTransitionGraph}
\end{algorithm}

For soundness, we have the following lemma. The proof, by induction on
the path length, is straightforward and is omitted.
\begin{lemma}
 \label{lemma:soundness}
  Let $M$ be an rSB model, and let $G$ be its transition graph
  constructed by Alg.~\ref{alg:spanTransitionGraph}. Then any
  execution path $q_0,q_1,\ldots$ of $M$, either finite or infinite,
  corresponds to a path $s_0,s_1,\ldots$ in $G$, and vice versa.
\end{lemma}

A natural concern is about the size of the power set, $2^{P_O}$. We
argue that this size is quite manageable: indeed, scenario objects
tend to be short and concise, as they deal only with specific aspects
of the system in question. Thus, spanning the individual transition
graphs should be doable, even for large systems. Of course, computing
the composite transition graph in Step 2 might suffer from the
infamous state explosion problem, which is common in
verification. Here, one possible solution is to break the model up
into sub-models, and reason about each of them separately; there exist
techniques for doing this~\cite{HaKaKa13}, but they are out of our
scope here.

\medskip
\noindent\textbf{Example.}
Observe the following pseudo-code, which represents Object 3 from
Fig.~\ref{fig:rsbmExample}:

\begin{lstlisting}
sync( ${\color{blue}request}=(v\geq 2 \vee h=0)$ );
sync( ${\color{blue}request}=(h\geq 10)$, ${\color{red}block}=(v\neq 0\vee h<0)$);
if ( $h<10$ )
  sync( ${\color{blue}request}=(h\geq 10)$, 
        ${\color{red}block}=(v\neq 0\vee h<10)$ );  
sync( ${\color{blue}request}=(true)$ );
\end{lstlisting}

In this case, syntactically constructing the set $P_O$ yields:
\[
  P_O=\{ v\geq 2, h=0, h\geq 10, v=0, h<0 \}
\]
(we do not need to consider a predicate and its negation, e.g.
$h\geq 10$ and $h<10$, twice). Thus, the power set $2^{P_O}$ has 32
elements. One of these elements is $\langle v \geq 2,
h=0\rangle$. Passing the constraint $v\geq 2\wedge h=0$ to an SMT
solver yields a concrete assignment, e.g.  $v=3,h=0$. Finally,
triggering this assignment in state $q_0$, which is the first
synchronization point of this scenario object, reveals a new state
which corresponds to the second synchronization point. However,
triggering the element $\{v=0\}$ does not result in reaching a new
state: the scenario object did not request this assignment, and so it
does not wake up from the synchronization call.

Repeating this process to saturation produces a transition graph
equivalent to the one depicted in Fig.~\ref{fig:rsbmExample}. (In
order to reduce clutter in the resulting graph, some transitions can
then be merged together by additional invocations of the SMT solver;
applying this approach is part of our ongoing work.)

\section{Formally Analyzing rSBM}
\label{sec:repairSbm}

\medskip
\noindent\textbf{Model checking rich SBM.}
Given Lemma~\ref{lemma:soundness}, it is straightforward to devise a
model-checking algorithm for rSBM models. Given a model $M$ and a
safety property $\varphi$, we can construct
the transition graph $G$ of $M$, compose it with $\varphi$, and then
check whether there are any reachable states that violate
$\varphi$. The rSBM formalism is sufficiently expressive to formulate
the property $\varphi$ itself as a scenario object that simply marks
some of its states as \emph{bad}. Then, the same techniques from
Section~\ref{sec:transitionGraphs} can be used to extract the
composite transition graph.

\begin{figure}[htp]
  \centering
  \scalebox{0.65} {
    
    \tikzstyle{box}=[draw,  text width=2.8cm,text centered,inner sep=3]
    \tikzstyle{set}=[text centered, text width = 10em]

    \begin{tikzpicture}[thick,auto,>=latex',line/.style ={draw, thick, -latex', shorten >=0pt}]
    
      \node (box1)  [box] {
        {\color{blue}Request} $\emptyset$\\
        {\color{red}Block} $\emptyset$
      };
      \node (box2)  [box, below=1cm of box1] {
        {\color{blue}Request} $\emptyset$\\
        {\color{red}Block} $\emptyset$
      };
      \node (box3)  [box, below=1cm of box2] {Bad};
      
      \path[->] (box1) edge [thick, bend left] node[] {$|v| \geq 4 \vee |h|\geq 18$} (box2);
      \path[->] (box2) edge [thick] node[] {$|v| \geq 4 \vee |h|\geq 18$} (box3);

      \path[] (box1) edge [loop right,thick] node {$|v|<4 \wedge |h| <
        18$} (box1);
      \path[->] (box2) edge [thick, bend left] node {$|v|<4 \wedge |h| < 18$} (box1);

      % \node (title) [box,draw=none] at ($(box1) + (-0.0cm,1.4cm)$) {\textsc{Object 1}};  
      % \node (title) [box,draw=none] at ($(box2) + (-0.0cm,1.4cm)$) {\textsc{Object 2}};  
      % \node (title) [box,draw=none] at ($(box3) + (-0.0cm,1.4cm)$) {\textsc{Object 3}};  
      
    \end{tikzpicture}
  }  
  \caption{Encoding a safety property as an rSBM scenario object.}
  \label{fig:badStates}
\end{figure}
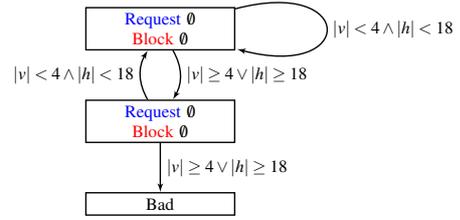

We demonstrate this process with an example. Let us review again the
model from Fig.~\ref{fig:rsbmExample}, and let us consider a safety
property stating that it is forbidden for the drone to turn sharply,
either vertically or horizontally, twice in a row. This property can
be encoded as the scenario object that appears in
Fig.~\ref{fig:badStates}. This object never requests or blocks
anything; it only waits for two consecutive events in which the
horizontal or vertical angular velocities were high ($|h|\geq 18$ or
$|v|\geq 4$, respectively). If such two consecutive events are
detected, the object marks the state as \emph{bad}.

By composing this new scenario object with the model depicted in
Fig.~\ref{fig:rsbmExample} and searching for reachable bad states
(e.g., using BFS), we can determine that a safety violation is
possible. Further, using the concrete variable assignments returned by
the SMT solver, we can provide a counter-example that demonstrates
this, e.g.: the triggering of $\langle v=5, h=0 \rangle$ and then of
$\langle v=0, h=20 \rangle$, which is allowed by the model.

\medskip
\noindent\textbf{Repairing rich SBM modes.}
Given an rSB model $M$ and a \emph{violated} safety property
$\varphi$, we now seek to repair  $M$ so that $\varphi$
becomes satisfied. Our goal is to cut off any bad states in the
underlying transition graph $G$, so that they become unreachable from
the initial state $q_0$. Further, we wish to cut off only paths that
lead, or are guaranteed to lead, to a violating state; that is, we do
not wish to remove any runs that are not violating.

To this end, we follow an approach proposed for repairing
vSBM~\cite{HaKaMaWe14} and utilize the blocking idiom to create
\emph{patch} scenario objects. Specifically, we propose to add to the
scenario-based model an additional scenario object that will apply
blocking at selected points during the execution, so as to cut off the
reachable bad states from the transition graph.  The algorithm for
performing this repair has the following steps:
\begin{inparaenum}[(i)]
\item given a behavioral model $M$ and a scenario object for a
  violated safety property $\varphi$, we extract the composite
  transition graph of $M$ and $\varphi$, as previously described. We
  then apply BFS to identify the set $B$ of all reachable bad states in this graph;
\item next, we iteratively search for states currently not in $B$,
  i.e. $q\notin B$, such that all of their outgoing edges lead to $B$,
  i.e. $q\to q'\Rightarrow q'\in B$. Any such state is
  \emph{guaranteed} to eventually lead to a violating state, and so is
  added to $B$, i.e. $B:=B\cup\{q\}$. This process is repeated until
  we reach a fixed point, and no new states can be added to $B$;
\item finally, we
  add a new scenario object to the model that keeps track of the
  execution, and blocks precisely those edges that lead to states in
  $B$. 
\end{inparaenum}
Intuitively, this process cuts off precisely those states that are
either bad themselves, or are guaranteed to lead to a bad state in a
finite number of steps. Thus, we only remove runs that violate the
safety property in question, and no others. Additionally, we create no
new bad runs, and do not introduce any deadlocks.

We demonstrate this process using the rSB model from
Fig.~\ref{fig:rsbmExample} and the violated safety property from
Fig.~\ref{fig:badStates}. The full, composite transition graph of this
model (including the safety property) is depicted in
Fig.~\ref{fig:compositeModel}. Unsurprisingly, there is a reachable
bad state in this model (state $q_6$). Our repair algorithm thus
starts from state $q_6$, i.e. $B=\{q_6\}$, and identifies all states
with edges leading to $q_6$; in this case, state $q_5$. Because state
$q_5$ has edges leading also to states not marked as bad (states
$q_3$ and $q_4$), it is not added to the set $B$ of bad states. Thus, the
algorithm generates a new scenario object such that, when composed
with the existing scenario objects, will apply blocking to prevent
state $q_5$ from transitioning into state $q_6$.

\begin{figure*}[htp]
  \centering
  \scalebox{0.65} {
    
    \tikzstyle{box}=[draw,  text width=6cm,text centered,inner sep=3]
    \tikzstyle{set}=[text centered, text width = 10em]

    \begin{tikzpicture}[thick,auto,>=latex',line/.style ={draw, thick, -latex', shorten >=0pt}]

      \node (box1)  [box, fill=white]
      {
        {\color{blue}Request} $v\geq 2 \wedge h=0$ \\
        {\color{red}Block} $h\leq -20 \vee h\geq 20 \vee v\leq -5
        \vee v\geq 5$
      };

      \node (box2)  [box, fill=white, below = 1cm of box1]
      {
        {\color{blue}Request} $h\geq 10$\\
        {\color{red}Block} $v\neq 0\vee h<0 \vee$\\
        $h\leq -20 \vee h\geq 20 \vee v\leq -5
        \vee v\geq 5$
      };
      \draw[->] (box1) -- node[] {$2\leq v< 4\wedge h=0$} (box2);

      \draw [->] ($(box1.north) + (0,0.5cm)$) -- (box1.north);

      \node (box3)  [box, fill=white, below = 1cm of box2]
      {
        {\color{blue}Request} $h\geq 10$\\
        {\color{red}Block} $v\neq 0\vee h<10 \vee$\\
        $h\leq -20 \vee h\geq 20 \vee v\leq -5
        \vee v\geq 5$
      };

      \draw[->] (box2) -- node[] {$h<10$} (box3);

      \node (box4)  [box, fill=white, below = 1cm of box3]
      {
        {\color{blue}Request} true\\
        {\color{red}Block} $h\leq -20 \vee h\geq 20 \vee v\leq -5
        \vee v\geq 5$
      };
%      \path[->] (box2) edge[->,transform canvas={xshift=-1.5cm}] node[swap] {$h\geq 10$} (box4);
      \path[->] (box2.west) edge[bend right] node[swap] {$h\geq 10$} (box4.west);

      \draw[->] (box3) -- node[] {$h\geq10$} (box4);
      
      \node (box5)  [box, fill=white] at ($(box2.east) + (6cm, 1.35cm)$)
      {
        {\color{blue}Request} $h\geq 10$\\
        {\color{red}Block} $v\neq 0\vee h<0 \vee$\\
        $h\leq -20 \vee h\geq 20 \vee v\leq -5
        \vee v\geq 5$
      };
      \path[->] (box1.east) edge [bend left] node[] {$v\geq 4\wedge h=0$} (box5.north);

      \path[->] (box5) edge [] node[pos=0.5,swap] {$10\leq h< 18$} (box4.north east);

      \path[->] (box5) edge [] node[pos=0.4,swap] {$h<10$} (box3);
      
      \path[] (box4) edge [loop below,thick] node {true} (box4);     

      \node (box6)  [box, below=3cm of box5] {Bad};

      \path[->] (box5) edge [] node[] {$h\geq 18$} (box6);

      \node (title) [box,draw=none,text width=0cm] at ($(box1.north
      west) + (-0.0cm,0.3cm)$) {$q_1$};
      \node (title) [box,draw=none,text width=0cm] at ($(box2.north
      west) + (-0.0cm,0.3cm)$) {$q_2$};
      \node (title) [box,draw=none,text width=0cm] at ($(box3.north
      west) + (-0.0cm,0.3cm)$) {$q_3$};
      \node (title) [box,draw=none,text width=0cm] at ($(box4.north
      west) + (-0.0cm,0.3cm)$) {$q_4$};

      \node (title) [box,draw=none,text width=0cm] at ($(box5.north
      east) + (-0.3cm,0.3cm)$) {$q_5$};
      \node (title) [box,draw=none,text width=0cm] at ($(box6.north
      east) + (-0.3cm,0.3cm)$) {$q_6$};

    \end{tikzpicture}
  }  
  \caption{The composite transition graph of the model from
    Fig.~\ref{fig:rsbmExample}, composed also with the scenario object
    for the safety property form Fig.~\ref{fig:badStates}.}
  \label{fig:compositeModel}
\end{figure*}

This new scenario object is depicted in Fig.~\ref{fig:patch}. It
merely waits for the sequence of events that would send the original
model into state $q_5$, namely an assignment that satisfies $v\geq
4\wedge h=0$; then blocks the assignments that would send the original
model into state $q_6$, namely $h\geq 18$; and then does nothing else
for the remainder of the run.

\begin{figure}[htp]
  \centering
  \scalebox{0.65} {
    
    \tikzstyle{box}=[draw,  text width=6cm,text centered,inner sep=3]
    \tikzstyle{set}=[text centered, text width = 10em]

    \begin{tikzpicture}[thick,auto,>=latex',line/.style ={draw, thick, -latex', shorten >=0pt}]

      \node (box1)  [box, fill=white]
      {
        {\color{blue}Request} $\emptyset$ \\
        {\color{red}Block} $\emptyset$
      };
      \draw [->] ($(box1.north) + (0,0.5cm)$) -- (box1.north);

      \node (box2)  [box, fill=white, below=1cm of box1]
      {
        {\color{blue}Request} $\emptyset$\\
        {\color{red}Block} $h\geq 18$
      };

      \draw[->] (box1) -- node[] {$v\geq 4\wedge h=0$} (box2);

      \node (box3)  [box, fill=white, below=1cm of box2]
      {
        {\color{blue}Request} $\emptyset$\\
        {\color{red}Block} $\emptyset$
      };

      \draw[->] (box2) -- node[] {$true$} (box3);

      \path[] (box3) edge [loop below,thick] node {true} (box3);
      
    \end{tikzpicture}
  }  
  \caption{The composite transition graph of the model from
    Fig.~\ref{fig:rsbmExample}, composed also with the scenario object
    for the safety property form Fig.~\ref{fig:badStates}.}
  \label{fig:patch}
\end{figure}

For soundness, we have the following lemma (proof omitted):
\begin{lemma}
  Let $M$ be an rSB model with reachable bad states, and let $M'$ be
  this model augmented with a patch scenario object, as explained
  above. The set of runs of $M'$ is then precisely the set of runs of
  $M$, with all violating runs removed.
\end{lemma}

We note that, while we have focused here on model checking and
repairing safety violations in rSB models, similar operations can be
performed also for \emph{liveness} properties.  Given a liveness
property $\varphi$, formulated as a scenario object that marks some
states as \emph{good}, we can check whether there exists a reachable
cycle in the composite rSB model that does not contain any good
states~\cite{BaKa08}. Further, if we detect such a cycle, a patch
scenario object can be added to the model to prevent it, again using
the \emph{blocking idiom}~\cite{HaKaMaWe14}. We leave treatment of
this case for future work.

\section{Related Work}
\label{sec:relatedWork}

The general research question that this paper addresses, namely how to
effectively model complex systems and then repair these models, has
been studied extensively. Here, we focused on the scenario-based
modeling paradigm, in which system behaviors are modeled as
scenarios~\cite{DaHa01,HaMa03,HaMaWe12ACM}. There are numerous related
approaches for modeling event-driven reactive systems: notable
examples include Esterel~\cite{BeGo92}, Lustre~\cite{HaCaRaPi91},
Signal~\cite{GuGaBoMa91}, and Petri Nets~\cite{HoKrGi97}. Similar
concepts appear also in component based programming languages, such as
\emph{BIP}~\cite{BaBoSi06}. Some of our repair techniques may be
carried over to these frameworks, provided that the \emph{blocking
  idiom}, which is crucial to our approach, is present or can be
achieved using other idioms. More broadly, scenario-based modeling is
adequate for modeling discrete event systems~\cite{CaLa09}; and the
repair of SB models is related to the \emph{supervisory control}
problem of such systems~\cite{RaWo87,RaWo89}.

Prior work has demonstrated how scenario-based models may be
automatically repaired~\cite{Ka13,HaKaMaWe14,HaKaMaWe12}, and here we
generalized and extended this approach to scenario-based models with
rich events --- which are necessary for modeling more complex
systems~\cite{KaMaSaWe19,Ka21b,ElMaStWe18,BaElSaWe19}. The automated
repair problem in other modeling and programming paradigms has also
been studied extensively: some examples include fault localization and
automatic repair by identifying sets of malfunctioning components and
synthesizing replacement
components~\cite{StJoBl05,StJoBl05b,JoGrBl05}; program repair via
semantic analysis~\cite{NgQiRoCh13}; the automatic repair of
concurrency-related bugs by analyzing execution traces associated with
bug reports~\cite{JiSoZhLuLi11}; applying genetic-programming to
repair legacy C programs~\cite{WeFoLeNg10}; combining
genetic-programming with co-evolution of test cases until a bug is
repaired~\cite{ArYa08}; and leveraging information regarding previous
repairs~\cite{LeLoLe16}. Naturally, work on automatic-repair of
models and programs can be considered a particular case of program
synthesis~\cite{PnRo89,BlJoPiPnSa12,AlBoJuMaRaSeSiSoToUd13}.

Scenario-based models have been automatically analyzed in a variety of
ways that go beyond repair. Notable examples include compositional
verification~\cite{HaKaKaMaMiWe13,HaKaMaWe15,HaKaLaMaWe15,KaBaHa15},
automated
optimization~\cite{GrGrKaMaGlGuKo16,StGrGrHaKaMa18},
synthesis~\cite{GrGrKaMa16}, and specification
mining~\cite{MaArElGoKaLaMaShSzWeHa16,HaKaMaMa16b}. It will be
interesting to extend these approaches to the rSBM setting, using the
techniques we have outlined here.

Our proposed approach relies on constraint solvers in order to
construct the model's underlying transition graph. The use of such
solvers in the context of software modeling is expending, with common
use-cases typically revolving around formal methods. Some examples
include \emph{symbolic execution}~\cite{PaVi09}, \emph{bounded
  model-checking}~\cite{BiCiClZh99}, and \emph{concolic
  testing}~\cite{Sen07}. The aforementioned techniques, and many
others, showcase the benefits that automated solvers afford in the
context of the various tasks that arise as part of a software model's
life cycle.

\section{Conclusion}
\label{sec:conclusion}

Scenario-based modeling is promising approach for designing and
implementing complex systems: on one hand, it is intuitive and
well-aligned with human perception of models, and on the other it is
compatible with automated analysis and repair of models. Our initial
results demonstrate that extending SBM to support rich events, which
may be required for modeling various real-world systems, does not harm
this compatibility: specifically, it is still possible to span the
underlying transition systems of models, and use these transition
systems for model-checking and automated repair.

We regard this paper as a first step in the direction of creating
automated analysis tools for rSBM. As part of our ongoing work we are
 pursuing several directions:
\begin{inparaenum}[(i)]
  \item implement our repair technique on top of an existing rSBM
    platform, and evaluate it on varied benchmarks;
  \item leverage our technique for spanning rSBM transition graphs to
    automate additional aspects of the system's life cycle, such as
    optimization~\cite{HaKaKaMaWeWi15}
    and specification mining~\cite{HaKaMaMa18}; and
  \item further improve the scalability of our technique for
    transition graph spanning.
  \end{inparaenum}
We hope that these lines of work will promote the use of rSBM in
additional systems and settings.

\section*{\uppercase{Acknowledgements}}

The project was partially supported by grants from the Binational
Science Foundation (2017662) and the Israel Science Foundation
(683/18).

\end{document}